\documentclass{siam-wns-article}
\usepackage{subfigure}
\usepackage{amsmath}
\usepackage{graphicx}
\usepackage{amssymb}
\title{Interdependent Network Formation Games}

\author{Juntao Chen and Quanyan Zhu}

\usehyperref 

\begin{document}
\maketitle

\section*{Summary}
Designing optimal interdependent networks is important for the robustness and efficiency of national critical infrastructures. Here, we establish a two-person game-theoretic model in which two network designers choose to maximize the global connectivity independently. This framework enables decentralized network design by using iterative algorithms. After a finite number of steps, the algorithm will converge to a Nash equilibrium, and yields the equilibrium topology of the network. We corroborate our results by using numerical experiments, and compare the Nash equilibrium solutions with their team solution counterparts. The experimental results of the game method and the team method provide design guidelines to increase the efficiency of the interdependent network formation games. Moreover, the proposed game framework can be generally applied to a diverse number of applications, including power system networks and international air transportation networks.

\section{Introduction}
Algebraic connectivity plays an important role on the speed of information spreading, synchronization and resilience to attacks in networks \cite{pinyu2014attact,martin2014algebraic}. Despite a significant number of works on network connectivity maximization, they have mainly focused on one single network. However, complex systems often consist of multiple networks each controlling themselves. The methodologies from previous studies may not be directly used to address such kind of network connectivity maximization problems. This motivates us to develop new tools for interdependent networks.

In this work, we use games to establish models and sign principles for interdependent networks. It has significant advantages over centralized methods when the network is large or composed of several interacting subnetworks. Our problem considers a global network $G$ composed of two interdependent networks, $G_1$ and $G_2$. The network designer of each subnetwork is seen as a player of the interdependent network formation games. Each player $P_i$, $i\in\{1,2\}$, can add or delete interlinks in $G_i$ and intra-links between $G_1$ and $G_2$, while subjecting to budget and link constraints. Adding a link to $G$ incurs a cost. $P_1$ and $P_2$ independently update the network in an alternating fashion by maximizing algebraic connectivity of $G$ at each step. At each round of the update, one player's optimal strategy is the best response to the strategy of the other player from the previous round. To be specific, we define $A_{i,k}$ and $D_{i,k}$ as the set of links that $P_i$ can add or delete at step $k$. $L_{{-i},k-1}$, where $-i\triangleq{\{1,2\}\setminus\{i\}}$, is the Laplacian matrix of $G$ obtained from the previous step after player $-i$ has chosen his network topology.  $x_e, x_e'\in\{0,1\}$ are decision variables of adding and removing links, respectively. Here, 1 means that link $e$ is selected into $G$, while 0 means otherwise. $x_{i,k}$ is a vector of length $|A_{i,k}|$ that consists of all $x_e$ and $x_{i,k}'$ is a vector of length $|D_{i,k}|$ that consists of all $x_e'$. $x_i$ is a column vector that denotes the state of all links that $P_i$ can form in $G$. $c_i$ is a vector with the same size of $x_i$ whose entries are the costs of corresponding links. $M_i$ and $k_i$ are the budget and link constraint for $P_i$, respectively. $\lambda_2(L)$ gives the second smallest eigenvalue of matrix $L$.
Then, the general problem for $P_i$ at step $k$ can be formulated as
\begin{equation}
\begin{split}\label{Q1}
\max \limits_{x_{i,k},\ x_{i,k}'}\ \lambda_2&(L_{-i,k-1}+\displaystyle\sum_{e=1}^{|A_{i,k}|}{x_e}{a_e}{a_e^T}-\displaystyle\sum_{e=1}^{|D_{i,k}|}{x_e'}{d_e}{d_e^T})\\
\qquad\mathrm{s.t.}\qquad&\displaystyle\sum_{e=1}^{|A_{i,k}|}{x_e}=\displaystyle\sum_{e=1}^{|D_{i,k}|}{x_e'},\\
&{c_i^T}{x_i}\le{M_i},\ \ \textbf{1}^{T}x_i=k_i,\\
&x_{i,k}\in\left\{{0,1}\right\}^{|A_{i,k}|},\ \ x_{i,k}'\in\left\{{0,1}\right\}^{|D_{i,k}|}.
\end{split}
\end{equation}

For comparison, we define a team problem where $P_1$ and $P_2$ cooperatively form the interdependent network.

\section{Main Results}
\begin{figure}[t]
  \centering
  \subfigure[Network model]{
    \label{greedy} 
    \includegraphics[width=1.7in]{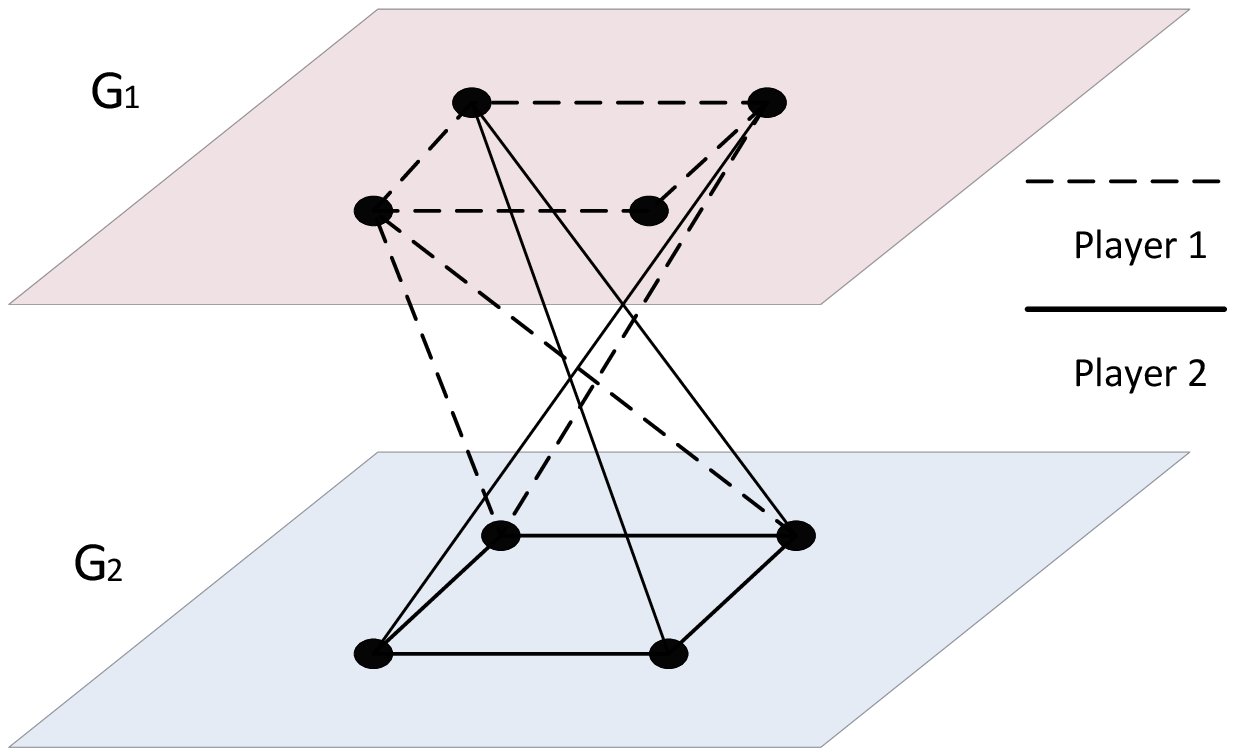}}
	 \subfigure[Performance of algorithms]{
    \label{rounding} 
    \includegraphics[width=1.5in]{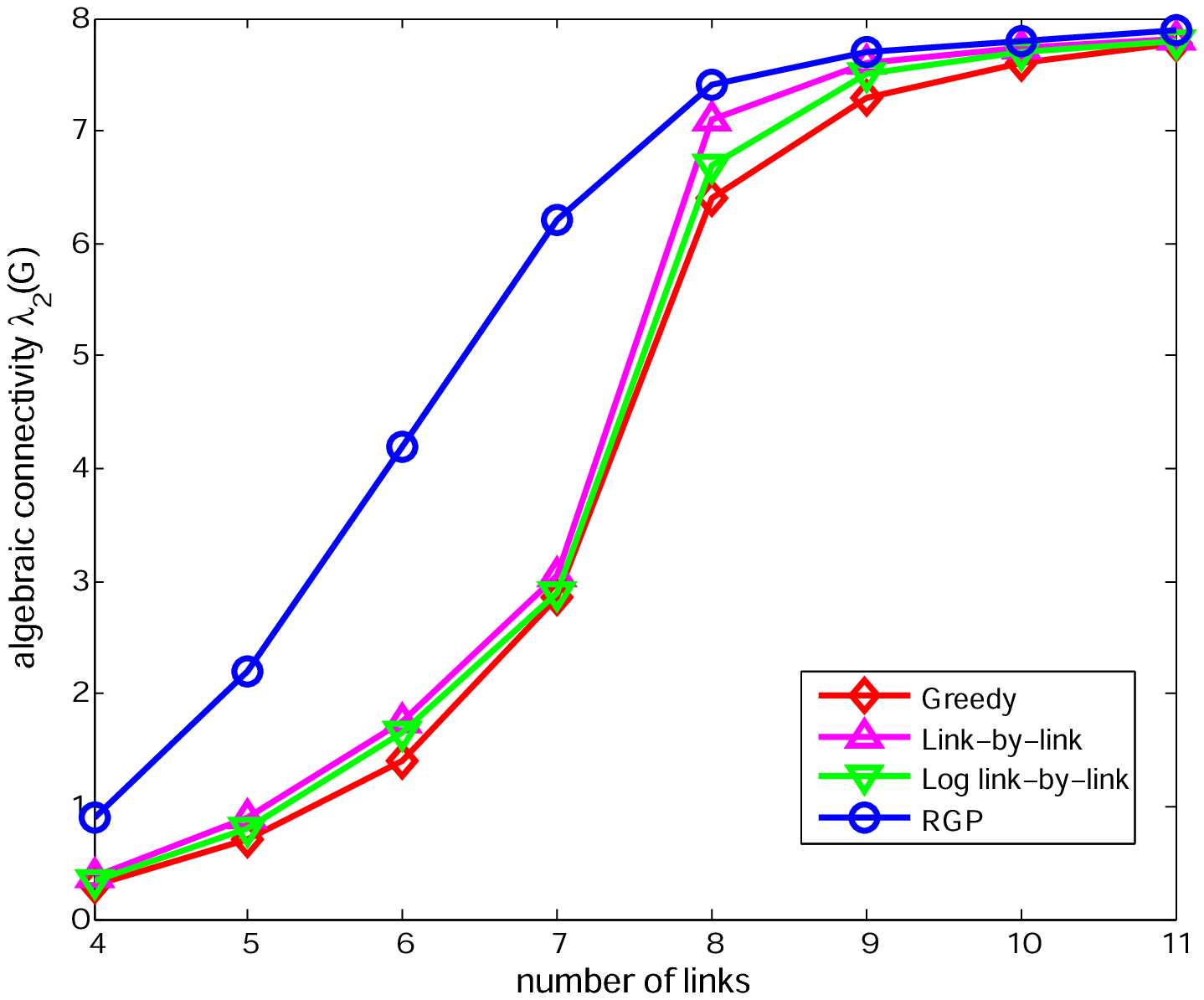}}
\caption{Connected links in (a) are the result of a case when $k_1=k_2=7$ and with greedy algorithm. Results of all steps in (b) are based on the model in (a) with $k_1=k_2$.}
  \label{f1} 
\end{figure}

Problem (1) can be reformulated into a simpler equivalent problem, where $P_i$ removes all links he has formed at previous step first, then starts from scratch and forms another $k_i$ links. Since each player chooses a set of links in the network, the game is a binary game problem (BGP). By relaxing the binary action set, we can transform BGP into a relaxed game problem (RGP), which is a semidefinite programming (SDP) problem. RGP is more convenient to solve than BGP. However, if the optimal solution of RGP is not feasible for BGP, then it should be rounded. Here, we consider three rounding techniques, i.e., greedy, link-by-link and log link-by-link. The difference between them is the number of elements chosen from the solution of RGP to round at each iteration. The performance of each algorithm of a case study is shown in Fig. \ref{f1}. $G_1$ and $G_2$ in Fig. \ref{greedy} are symmetric, and both contain 4 nodes. Clearly, $\lambda_2$ of RGP is an upper bound for the optimal solution of BGP with rounding. In terms of computation time and optimality, greedy algorithm yields the best result. All curves in Fig. \ref{rounding} are lower bounded by 8 corresponding to the completely connected network.

\begin{figure}[t]
  \centering
  \subfigure[Power system networks]{
    \label{simulationmodel} 
    \includegraphics[width=1.60in]{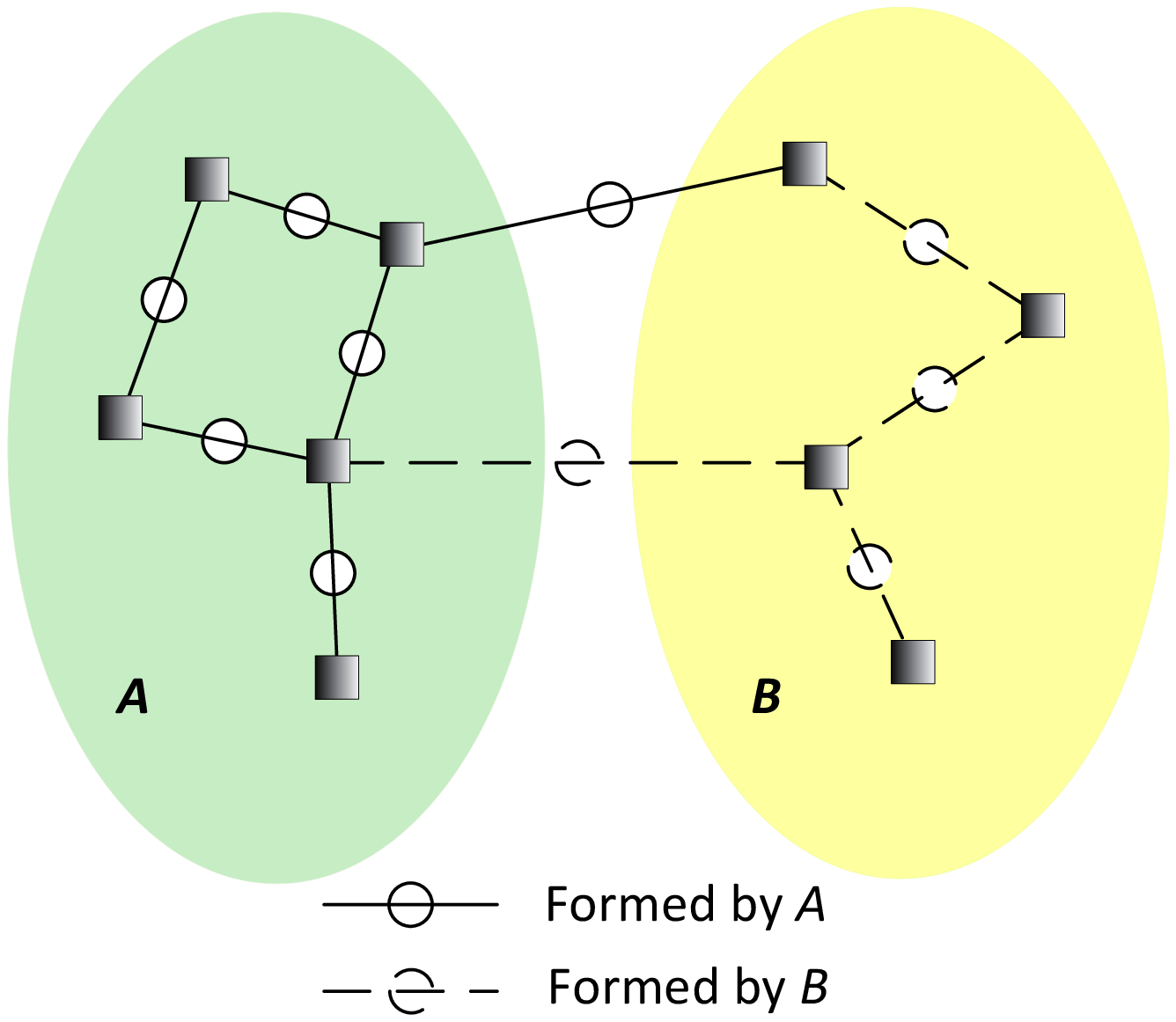}}
	 \subfigure[Performance of GM and TM]{
    \label{example} 
    \includegraphics[width=1.60in]{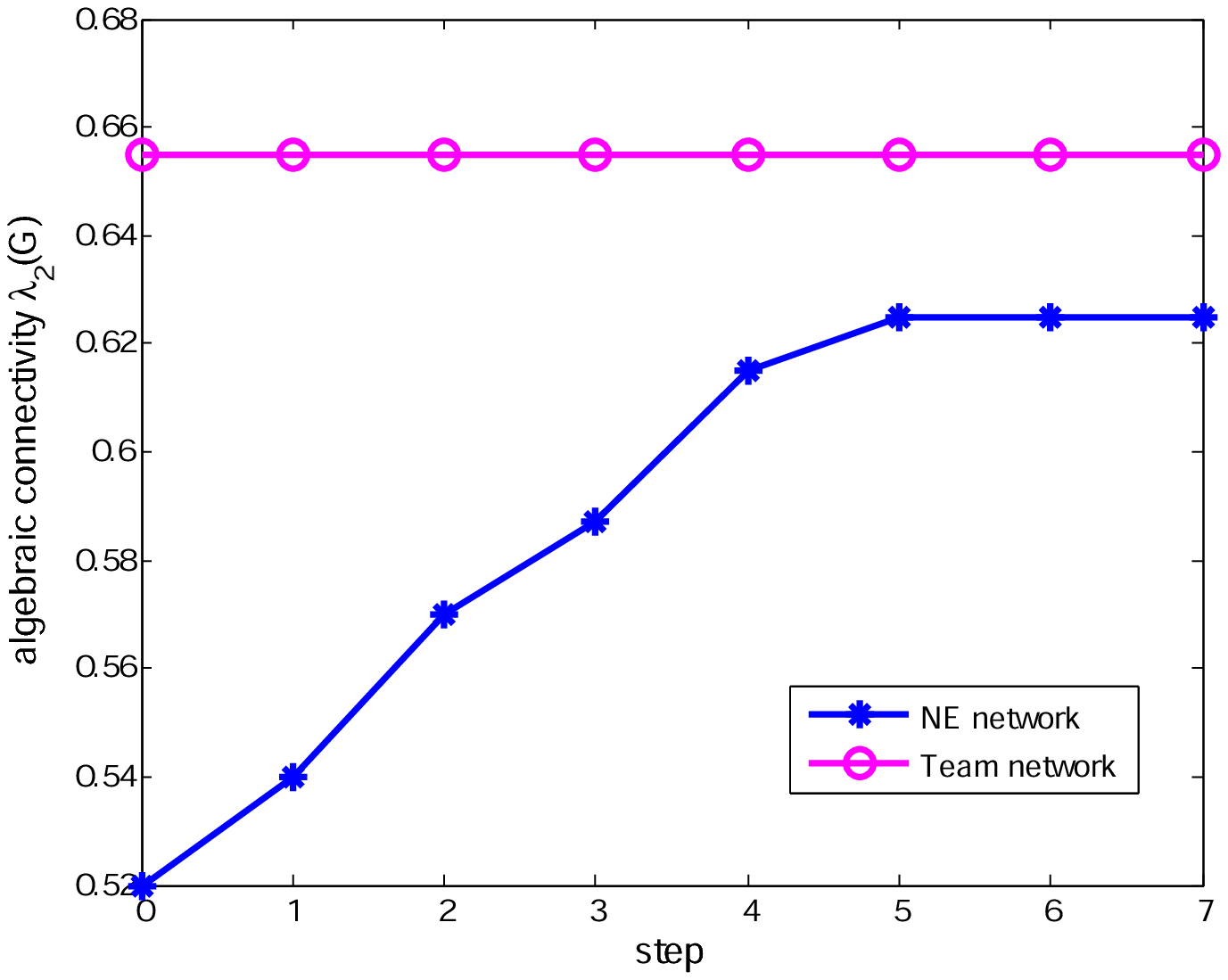}}
\caption{(a) gives a model of two PSNs. (b) shows the performance of GM and TM based on the PSN in (a).}
  \label{f2} 
\end{figure}
We compare the performance of game model (GM) and team model (TM) via a numerical example of interdependent power system networks (PSN). The network model is shown in Fig. \ref{simulationmodel}. $A$ and $B$ have 5 and 4 substations, respectively, and $k_1=6$ and $k_2=4$ in the existing network. Each power line (PL) has a cost related to the power transmission loss. Based on this network, both game model and team model are used to design the PSN. As shown in Fig. \ref{example}, the connectivity of team network is better than that of the Nash equilibrium (NE) network, and after 5 steps, NE network reaches an equilibrium. Loss of connectivity (LOC), which measures the gap between the two solutions, is 4.6\%.

\begin{figure}[t]
  \centering
  \subfigure[]{
    \label{limpact} 
    \includegraphics[width=1.5in]{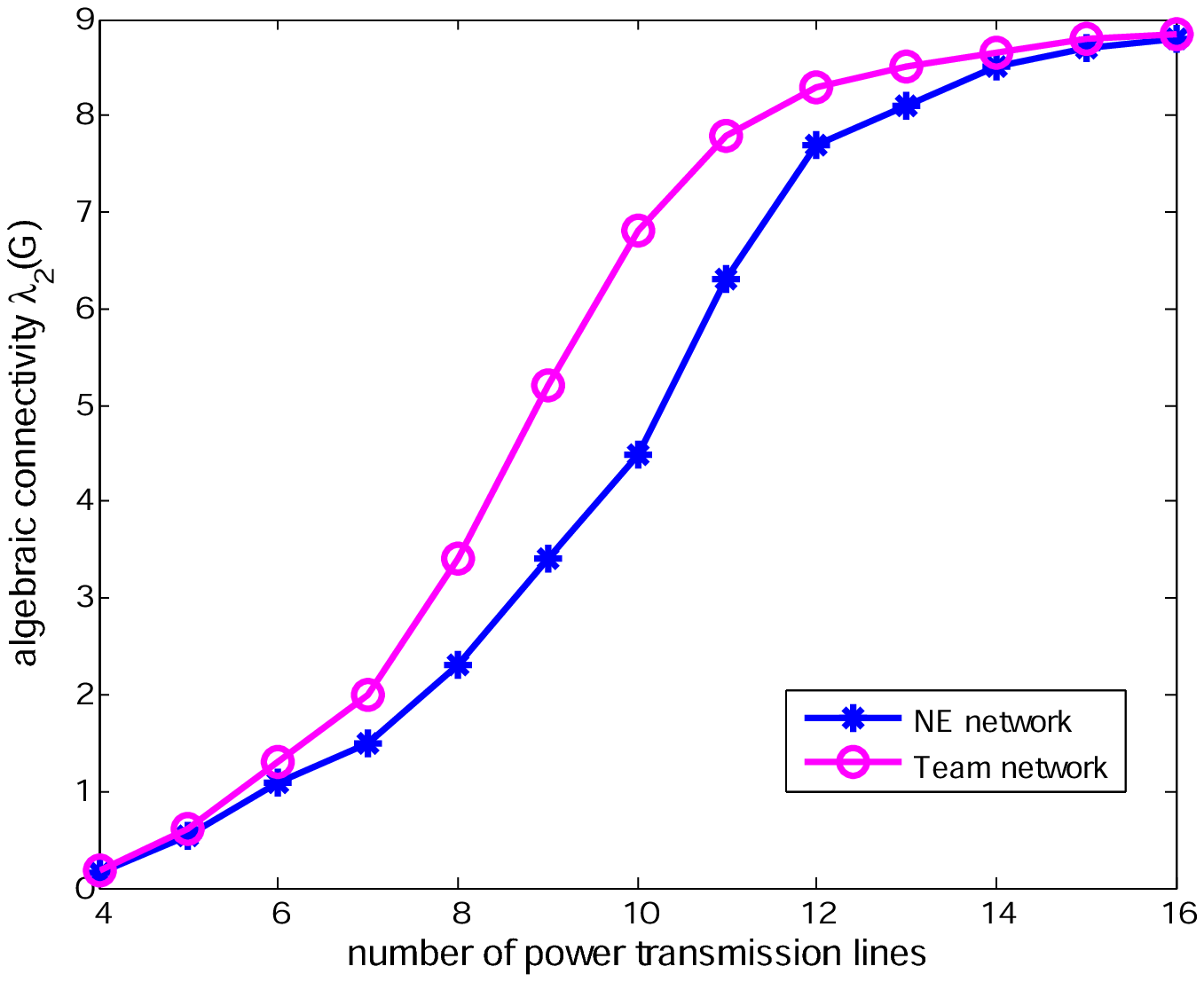}}
	 \subfigure[]{
    \label{nimpact} 
    \includegraphics[width=1.5in]{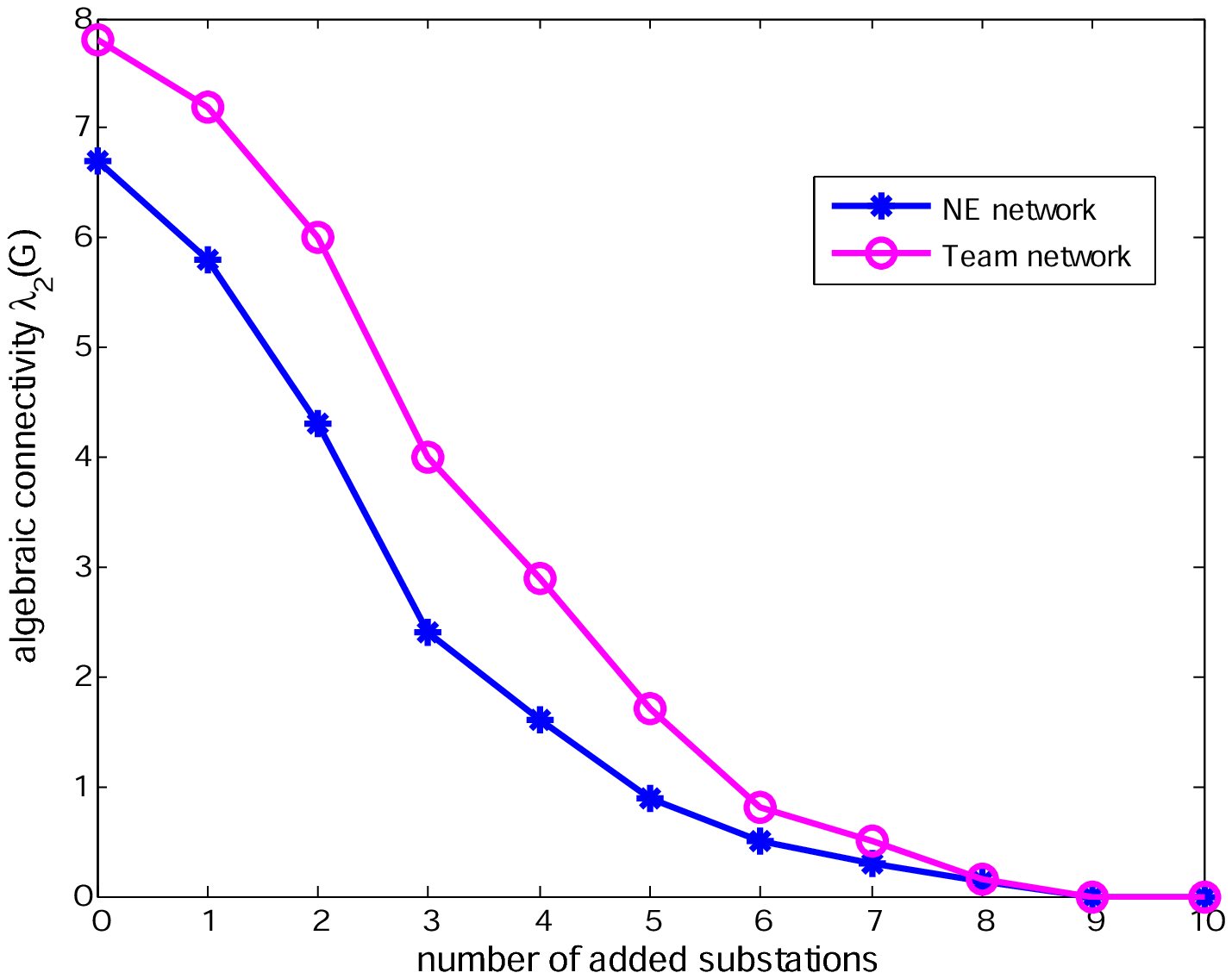}}
\caption{(a) and (b) are the impact of number of PL and substation on network connectivity respectively.}
  \label{f3} 
\end{figure}

To obtain more general results, we focus on the impact of the number of PL and substations, respectively. In the first case, we relax the constraint on budget, and assume $k_1=k_2$. In the second case, the budget is fixed, and the number of substations in both networks is increased by one at every step. We assume the cost of each PL is the same (ideal PL), which leads to fixed $k_1$ and $k_2$. Simulation results of these two cases are shown in Fig. \ref{f3}. At $k_1=9,10,11$ in Fig. \ref{limpact}, LOCs are equal to 32.3\%, 34.8\% and 33.8\%, respectively, higher than those of other steps. Therefore, the team solution has a better performance over the game solution in these cases. However, when $k_1$ and $k_2$ are greater than 14 or smaller than 7 in this case, there is no sizable difference between noncooperative and cooperative PSN formations. GM is preferable for these two regions in terms of reduced complexity and information security since players need not to share budget information. Fig. \ref{nimpact} provides design insights that cooperative PSN is better when a moderate number of substations are added to the network.

\bibliographystyle{abbrv}
\bibliography{wns-bib}

\end{document}